\documentclass[sigconf]{acmart}

\usepackage{amsmath,amsfonts,amsthm}
\usepackage{graphicx}
\usepackage{textcomp}
\usepackage[mathscr]{eucal}
\usepackage{multirow}
\usepackage{setspace}
\usepackage{xcolor,colortbl,framed}
\usepackage{tabularx}
\usepackage{float}
\usepackage{url}
\usepackage{enumitem}
\usepackage{balance}
\usepackage{booktabs}
\usepackage{threeparttable}
\usepackage{pifont}
\usepackage{bbding}

\usepackage[most]{tcolorbox}

\definecolor{myred}{RGB}{208,31,47}
\usepackage{tikz}

\newcommand{\redcircled}[1]{%
    \tikz[baseline=(char.base)]{%
        \node[shape=circle,fill=myred,text=white,inner sep=0.1ex] (char) {#1};%
    }%
}

\newcommand{\code}[1]{{\fontfamily{cmtt}\fontseries{m}\fontshape{n}\selectfont\small{#1}}}

\usepackage{xspace}
\newcommand{\T}{\textsc{MAS-SZZ}\xspace}

\AtBeginDocument{%
  \providecommand\BibTeX{{%
    \normalfont B\kern-0.5em{\scshape i\kern-0.25em b}\kern-0.8em\TeX}}}

\begin{document}
\newcolumntype{C}[1]{>{\centering\arraybackslash}p{#1}}
\title{\T: Multi-Agentic SZZ Algorithm for Vulnerability-Inducing Commit Identification}


\author{Sicong Cao$^\dagger$, Jinxuan Xu$^\dagger$, Le Yu$^\dagger$$^\clubsuit$, Jing Yang$^\ddagger$, Xingwei Lin$^\ddagger$, Linlin Zhu$^\S$, Fu Xiao$^\dagger$}

\affiliation{%
  \institution{%
    $^\dagger$Nanjing University of Posts and Telecommunications \quad
    $^\ddagger$Zhejiang University \\
    $^\S$DBAPPSecurity Co., Ltd. \quad
    $^\clubsuit$Corresponding author
  }
  \country{}
}

\email{{sicong.cao, 1225045415, yulele08, xiaof}@njupt.edu.cn}
\email{{22521073, xwlin.roy}@zju.edu.cn, jennie.zhu@dbappsecurity.com.cn}

\renewcommand{\shortauthors}{Cao et al.}

\begin{abstract}
Accurate vulnerability-inducing commit identification serves as a foundation for a series of software security tasks, such as vulnerability detection and affected version analysis. A straightforward solution is the SZZ algorithm, which traces back through the code history to identify the earliest commit that modify the vulnerable code. Unfortunately, neither the customized V-SZZ nor state-of-the-art LLM4SZZ perform satisfactorily due to the incorrect anchor selection and inadequate backtracking capability, making them far beyond a reliable usage in practice.

To overcome these challenges, we propose a multi-agentic SZZ algorithm, named \T, that facilitates the identification of vulnerability-inducing commits through collaboration among agents. Specifically, given a CVE description and its corresponding fixing commit, \T summarizes the root cause of the vulnerability and employs a structured step-forward prompting strategy to localize vulnerability-related statements based on the change intent of each patch hunk. These vulnerable statements serve as anchors from which \T autonomously traces backward through the repository's history to find the commit that first introduced the vulnerability. Extensive experiments show that \T outperforms the state-of-the-art baselines across datasets and programming languages, achieving F1-score gains of up to 65.22\% over the best-performing SZZ algorithm.
\end{abstract}


\maketitle

\section{Introduction}
Unpatched software versions with 1-day vulnerabilities pose severe threats to end users, as adversaries could learn from released patches to find and exploit them, incurring massive remediation and operational costs~\cite{DBLP:conf/issta/YangHCMLXCZ23}.
Therefore, it is essential to determine which versions are affected by a known vulnerability, underpinning a wide range of downstream tasks such as vulnerability detection~\cite{DBLP:journals/tdsc/SunZCWBWLX25} and severity assessment~\cite{DBLP:conf/sigsoft/LiYZWN23}. A straightforward solution is the SZZ algorithm~\cite{bszz,agszz,LRSZZ,MASZZ}, which identifies the \textbf{Vulnerability-Inducing Commits (VICs)} that introduced the security-critical bugs, from the \textbf{Vulnerability-Fixing Commits (VFCs)}. The pioneering work V-SZZ~\cite{bao2022v} leverages the line mapping algorithms to identify the earliest commit that modified the vulnerable lines as VICs.

Despite receiving great attention from academics and practitioners, neither the foundational SZZ algorithm nor its improved variants perform satisfactorily. Even though the state-of-the-art approach LLM4SZZ \cite{LLM4SZZ} achieves an accuracy of only 40.7\% on the manually curated dataset \cite{DBLP:conf/kbse/ChenLCXCLSSCH25}, which is far beyond a reliable usage in practice. Such poor ability stems from two major \textbf{challenges}:

\noindent\textbf{Challenge 1: Incorrect anchor selection.}
Based on the core assumption that only deleted or modified lines cause vulnerabilities, most SZZ algorithms adopt simple diff-based heuristics to select anchor statements for backtracking. Nonetheless, due to the existence of composite \cite{DBLP:journals/ese/HerzigJZ16} and ghost \cite{DBLP:journals/tse/RezkKM22} commits, they often fail to localize the true root cause line, resulting in high false-positive rate. Although LLM4SZZ tries to address this issue using Large Language Model (LLM) for contextual filtering, it achieves limited success as vulnerability-inducing statements are sometimes hidden in the unmodified context.

\noindent\textbf{Challenge 2: Inadequate backtracking capability.}
Standard SZZ algorithms apply one-step backtracking, which often misses earlier commits that introduced the vulnerability. V-SZZ is the first to introduce the multiple backtracking capability, which performs consecutive tracing until it identifies a commit as a stopping point. However, it still suffers from the over-traced or under-traced issue because of the use of similarity-based heuristics.

\noindent\textbf{Our work.}
In this paper, we present \T, a novel multi-agentic SZZ algorithm for VIC identification. Specifically, given a CVE description and its corresponding VFC, \T first summarizes the root cause of the vulnerability, accompanied by verified evidence points that trace each claim back to the supplied inputs. Then, \T employs a structured step-forward prompting strategy to infer the change intent of each patch hunk, and identifies the root cause lines from vulnerability-related hunks for backtracking, addressing \emph{Challenge 1}. Third, starting from an anchor statement, \T iteratively backtracking the commit history guided by the root cause, and dynamically determines the earliest vulnerability-contained commit as the VIC, resolving \emph{Challenge 2}.

\noindent\textbf{Evaluation.}
We implement a prototype system of \T, and conduct comparative experiments with seven representative SZZ algorithms \cite{bszz,agszz,MASZZ,bao2022v,LRSZZ,LLM4SZZ} on the two well-regarded benchmark datasets \cite{bao2022v,DBLP:journals/tosem/JavaSZZ}. Experimental results show that \T outperforms the state-of-the-art baselines with respect to F1-score across datasets and programming languages. In particular, \T achieves substantial gains, improving F1-score by up to 65.22\% over the best-performing baseline.

\noindent\textbf{Contributions.}
This paper makes the following contributions:
\begin{itemize}[leftmargin=1em]
    \item We design and implement \T, a multi-agent system-based SZZ algorithm for VIC identification.

    \item We propose several optimizations, including evidence-grounded root cause analysis, intent-driven anchor statement selection, and autonomous repository exploration, to alleviate the limitations of existing SZZ algorithms.
    
    \item We evaluate \T against seven existing SZZ algorithms on the two human-annotated datasets. The extensive experiment results indicate that \T outperforms the state-of-the-art baselines across programming languages.
\end{itemize}

\noindent\textbf{Paper Structure.}
Section~\ref{background} introduces the background on SZZ algorithms and LLM-based agents. Section~\ref{methodology} describes the details about our approach. Section~\ref{sec:experiments} presents the experimental setup, followed by the evaluation results. Section~\ref{sec:conclude} concludes the paper.

\section{Background and Related Work}\label{background}
\noindent\textbf{SZZ Algorithms.}
The initial SZZ algorithm, B-SZZ~\cite{bszz}, is based on the premise that ``\emph{a bug-inducing commit adds lines that are later removed by a fix''}. It traces deleted or modified lines back to their most recent commit. Over the past two decades, numerous variants \cite{agszz,LRSZZ,MASZZ,TCSZZ,LLM4SZZ} have been proposed to improve performance. AG-SZZ \cite{agszz} filters out cosmetic changes such as whitespace and comments based on annotation graphs to significantly reducing false positives, while MA-SZZ \cite{MASZZ} extends this by excluding meta-changes (e.g., branch merges and property changes) that do not alter program behavior. Recently, LLM4SZZ \cite{LLM4SZZ} further leverages LLMs to assess blame candidates with expanded context, achieving state-of-the-art results across multiple datasets.

As an early attempt to extend SZZ algorithms for VIC identification, Nguyen et al.~\cite{nguyen2016automatic} applied B-SZZ, serving as a foundational starting point for subsequent research.
Bao et al.~\cite{bao2022v} later proposed V-SZZ, an improved approach that labels the earliest commit as the VIC by computing line similarity.

\noindent\textbf{LLM-Based Agents.}
LLM agents, hereinafter also referred to as \emph{agents} for short, involve LLM applications that can execute complex tasks through the use of an architecture that combines LLMs with key modules like planning and memory. When building agents, an LLM serves as the main controller that controls a flow of operations needed to complete a task or user request. In recent years, a number of agent-driven solutions have been proposed to solve real-world software problems such as vulnerability detection \cite{DBLP:journals/tosem/ZhouCSL25} and GitHub issue resolution \cite{DBLP:conf/nips/0003ZWZ0C24} . For example, Claude Code\footnote{\url{https://github.com/anthropics/claude-code}} is a widely adopted proprietary coding agent, while SWE-agent \cite{DBLP:conf/nips/YangJWLYNP24} bridges LLMs and terminal environments by structuring tool interactions.

\section{Methodology}\label{methodology}
\begin{figure}[t]
\centering
\includegraphics[width=\linewidth]{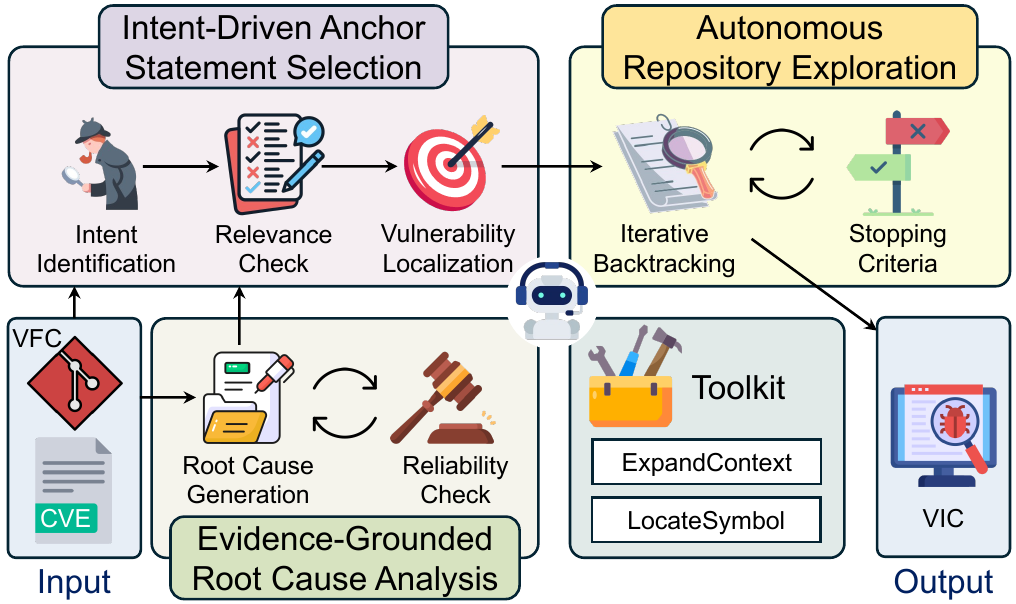}
\caption{Overview of \T.}
\vspace{-1.1em}
\label{fig:pipeline}
\end{figure}

\subsection{Overview}
As shown in Figure~\ref{fig:pipeline}, \T operates at three main stages: 

\begin{itemize}[leftmargin=1em]
    \item \textbf{Evidence-Grounded Root Cause Analysis:} Given a CVE description and its corresponding VFC (commit message \& code diff), the \emph{Auditor} agent summarizes the root cause of the vulnerability, accompanied by evidence points that trace each claim back to the supplied inputs. The root cause, along with the evidence, is then provided to the \emph{Judge} agent for further confirmation.
    \item \textbf{Intent-Driven Anchor Statement Selection:} The \emph{Reviewer} agent identifies the change intent of each patch hunk via a structured step-forward prompting strategy, and filters out vulnerability-irrelevant ones based on the root cause. Then, the \emph{Locator} agent jointly analyzes each changed statement and its surrounding context to identify the root cause line as anchor for backtracking.
    \item \textbf{Autonomous Repository Exploration:} Starting from an anchor statement, \T walks the commit history backward, invoking a \emph{Tracer} agent guided by the root cause at each step to determine whether the vulnerability exists in that commit, terminating when the vulnerability is no longer found and returning the last confirmed vulnerability-contained commit as the VIC.
\end{itemize}

\subsection{Evidence-Grounded Root Cause Analysis}
Directly querying LLMs to identify vulnerability-related statements may suffer from the hallucination issue \cite{DBLP:journals/tse/ZhangWLSZML26}, leading to incorrect backtracking judgments. Instead, \T produces a grounded explanation of why the vulnerability exists, which serve as a reference that guides both anchor statement selection and vulnerability presence judgments throughout the commit backtracking process.

Specifically, given a CVE description and the hash of its corresponding VFC, \T retrieves the commit message and diff (five lines of context as default) via \code{git show}. Each of them contributes a distinct but complementary perspective: CVE description serves as a summary of the vulnerability and can include information such as the impacts, attack vector, or other relevant technical information, commit message conveys the developer's intent behind the change, and diff exposes the code-level evidence of what was wrong. Based on these information, the \emph{Auditor} agent summarizes the root cause of the vulnerability, accompanied by evidence points that trace each claim back to the supplied inputs.

Once a root cause is output, the \emph{Judge} agent checks its reliability from two dimensions: \textbf{Evidence Traceability} and \textbf{Logical Consistency}. The former measures whether each evidence point can be associated with specific patch hunk in the diff or CVE description, the latter evaluates whether the stated root cause is logically consistent with the commit message. When either criterion fails, the \emph{Judge} agent formulates the corrective feedback that identifies precisely what is missing or incorrect. This feedback is passed to the \emph{Auditor} agent as supplementary information on the next attempt, so as to pargetted addresses the identified shortcomings rather than re-generate blindly. This generate-and-review loop will not terminate until the given attempt budget  (three rounds by default) expires or a root cause is marked with a \code{Pass} decision.

\subsection{Intent-Driven Anchor Statement Selection}
Given the fact that VFCs are sometimes entangled with non-security changes~\cite{DBLP:conf/uss/SunX00L025}, such as code refactoring and new features, treating all patch hunks uniformly results in the inclusion of noisy statements as candidates, which weakens the signal used for backtracking. Thus, in this stage, \T infers the intent of each patch hunk, and identifies root cause lines from vulnerability-related patch hunks as anchor statements for backtracking.

\noindent\textbf{Intent Identification.}
Before inferring the change intent, the \emph{Reviewer} agent first gather necessary information on demand from the codebase to understand the semantics of each patch hunk. To support this process, \T provides two tools called \emph{ExpandContext} and \emph{LocateSymbol}:
\begin{itemize}[leftmargin=1em]
    \item \emph{\textbf{ExpandContext.}} This tool specifies the range of code lines to be viewed in current commit, and returns the code snippet between the specified line numbers, enabling the \emph{Reviewer} agent to access the broader context.
    
    \item \emph{\textbf{LocateSymbol.}} This tool allows searching for file contents in git repository files. It takes simple string and greps the entire codebase, enabling the \emph{Reviewer} agent to efficiently localize required keywords, functions, or variables.
\end{itemize}

Incorporating the expanded context, the \emph{Reviewer} agent employs a step-forward prompting strategy \cite{DBLP:journals/corr/abs-2601-01233}, which enforces a four-step Chain-of-Thought (CoT) \cite{DBLP:conf/nips/Wei0SBIXCLZ22} sequence that mirrors how an expert developer analyzes a diff to identify the intent behind each patch hunk. \textbf{First}, the agent is prompted to describe the code-level modifications, anchoring the analysis to observable evidence rather than speculation. \textbf{Second}, it reasons about how that modification affects program behavior at runtime. \textbf{Third}, the agent infers the developer's intent and classifies each patch hunk into one of the ten distinct categories, such as ``fix'', ``chore'', and ``refactor'', according to the Conventional Commits Specification (CCS) \cite{DBLP:conf/icse/ZengZQL25}. \textbf{Fourth}, the agent distills the entire analytical chain into a structured 2-tuple $\langle$Change Category, Intent Summary$\rangle$.

\noindent\textbf{Relevance Check.}
After inferring the intent of each patch hunk, the \emph{Evaluator} agent further conducts the relevance check to filter out vulnerability-irrelevant parts of a VFC. Particularly, given a candidate hunk (i.e., change category is marked as ``fix''), the \emph{Evaluator} agent assesses whether its individual intent aligns logically with the root cause summarized before, and returns a binary \code{RELEVANT} or \code{IRRELEVANT} decision.

\noindent\textbf{Vulnerability Localization.}
For each hunk labeled as \code{RELEVANT}, the \emph{Locator} agent selects the specific statement(s) most directly responsible for the vulnerability based on (\ding{182}) the code changes, (\ding{183}) their surrounding context, and (\ding{184}) the summarized root cause.

\subsection{Autonomous Repository Exploration}

The localized vulnerable statements serve as anchors from which \T traces backward through the repository's history to find the commit that first introduced the vulnerability.

\noindent\textbf{Iterative Backtracking.} For each anchor statement, the \emph{Tracer} agent begins at the immediately previous commit to the lines changed in the VFC, and iteratively executes the \code{git blame} command to go back to identify the descendants commits. For the descendants commit retrieved in each round, the \emph{Tracer} agent analyzes its commit message and the post-patch code to determine whether the vulnerability was present at that revision. In this phase, the \emph{Tracer} agent is also quipped with two customized tools we mentioned before, i.e., \emph{ExpandContext} and \emph{LocateSymbol}, to obtain the necessary contextual information for decision-making.

\noindent\textbf{Stopping Criteria.}
Unlike existing SZZ algorithms that terminate after a fixed number of iterations, ranging from a single step~\cite{bszz} to full history traversal~\cite{bao2022v}, the \emph{Tracer} agent dynamically determines when to stop by comparing the pre- and post-patch version. Particularly, when the vulnerability is no longer found in current commit but still exists in the previous ones, the \emph{Tracer} agent terminates and returns the last vulnerability-contained commit as the VIC.

\subsection{Running Example}
\begin{figure*}[ht]
\centering
\includegraphics[width=.96\textwidth]{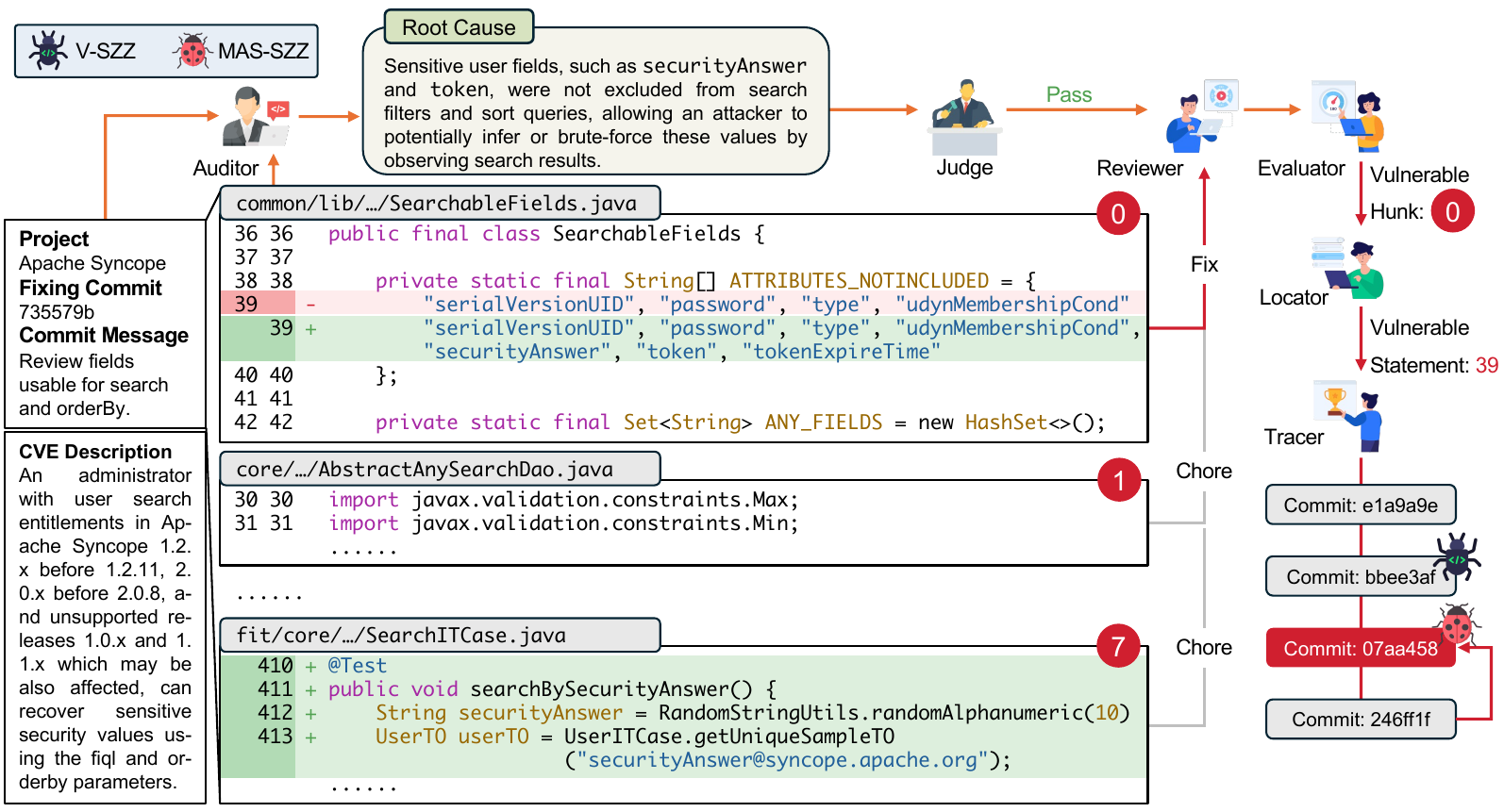}
\caption{A running example detailing the full workflow of \T on CVE-2018-1322.}
\vspace{-1.1em}
\label{fig:example}
\end{figure*}

To better illustrate how \T works, we take CVE-2018-1322\footnote{\url{https://github.com/apache/syncope/commit/735579b6f987b407049ac1f1da08e675d957c3e}} as an example. By default, Apache Syncope prevents sensitive values from being returned when querying the \code{/syncope/rest/users} endpoint. Fields such as \code{securityAnswers} or \code{password} will always return \code{null}. However the results returned can be filtered or sorted based on sensitive fields. By measuring how the results are returned the values of the desired fields can be successfully recovered. The \code{fiql} parameter can be used to recover full security answers, and the \code{orderby} parameter can be used to recover full security answers and partial information about password hashes. As shown in Figure \ref{fig:example}, the fixing commit \code{735579b} of this vulnerability spans five files, while the true root cause is only concentrated in a single statement: line 39 in \code{SearchableFields.java}. According to the manual annotation, the commit \code{07aa458} is labeled as the associated VIC.

\noindent\textbf{Stage 1: Root Cause Analysis.}
The \emph{Auditor} agent first analyzes the description, commit message, and code diff of this vulnerability, and summarizes the root cause (the evidence points are omitted due to the limited space) as follows: \emph{Sensitive user fields, such as \code{securityAnswer} and \code{token}, were not excluded from search filters and sort queries, allowing an attacker to potentially infer or brute-force these values by observing search results.} Then, the \emph{Judge} agent confirms that (\ding{182}) each evidence point can be associated with specific patch hunk in the diff or CVE description, and (\ding{183}) the stated root cause is logically consistent with the commit message, delivering them to the next stage. 

\noindent\textbf{Stage 2: Anchor Statement Selection.}
The \emph{Reviewer} agent infers the change intent of each patch hunk. For example, hunk \redcircled{7} in \code{SearchITCase.java} adds a JUnit annotation to test whether a user can be located by searching with a plain‑text security answer. It is classified as ``chore'' and discarded by the \emph{Reviewer} agent since this hunk aims to verify the completeness of the patch in a post-hoc manner, rather than directly fix the vulnerability. By contrast, hunk \redcircled{0}, which appends \code{securityAnswer}, \code{token}, and \code{tokenExpireTime} in \code{SearchableFields.java} to \code{ATTRIBUTES\_NOTINCLUDED}, is classified as ``fix'' and retained. All hunks labeled as ``fix'' are fed into the \emph{Evaluator} agent for relevance check. Finally, one out of eight patch hunks (i.e., hunk \redcircled{0}) is considered to be vulnerability-related. In conjunction with the established root cause, the \emph{Locator} agent subsequently selects line 39 as the anchor statement for backtracking.

\noindent\textbf{Stage 3: Autonomous Repository Exploration.}
Starting from line 39, the \emph{Tracer} agent
backtracks iteratively via \code{git blame}:
\begin{itemize}[leftmargin=1em]
    \item \textbf{Commit \code{e1a9a9e}:} The commit removes \code{realm} from the array, but \code{securityAnswer}, \code{token}, and \code{tokenExpireTime} remain absent. The vulnerability is judged present and backtracking continues.

    \item \textbf{Commit \code{bbee3af}:} The commit expands the array from two fields to five, but the three sensitive fields are still missing. As a result, the vulnerability is judged present and backtracking continues. It's noteworthy that the pioneering V-SZZ stops the traceback here since the line similarity is far less than 0.75.

    \item \textbf{Commit \code{07aa458}:} This commit first introduces \code{serialVersionUID} and \code{password}, but the sensitive fields are never considered. Thus, the vulnerability is judged present and backtracking continues.

    \item \textbf{Commit \code{246ff1f}:} Class \code{SearchableFields} does not yet exist at this revision. The \emph{Tracer} agent judges the vulnerability absent, and returns the previous commit \code{07aa458} as the VIC.
\end{itemize}

\section{Evaluation}\label{sec:experiments}
\subsection{Research Questions}
Our work seeks to answer three \textbf{Research Questions (RQs)}:
\begin{itemize}[leftmargin=1em]
    \item \textbf{RQ1:} Which LLM backbone is the best for \T?

    \item \textbf{RQ2:} How does \T perform in comparison to state-of-the-art SZZ algorithms?
    
    \item \textbf{RQ3:} How do various components of \T affect its overall performance?

\end{itemize}

\subsection{Experimental Setting}
\noindent\textbf{Datasets.}
We employ two manually-validated datasets, including V-SZZ~\cite{bao2022v} and Java-SZZ~\cite{DBLP:journals/tosem/JavaSZZ}, for evaluation. The \textbf{V-SZZ} dataset contains 100 C/C++ and 69 (out of 72\footnote{Three Java cases were discarded due to the lack of valid VICs (CVE-2013-0239 and CVE-2018-17192) or available repositories (blynk-server).}) Java cases, while the \textbf{Java-SZZ} dataset comprises 100 Java VICs.

\noindent\textbf{Baselines.}
Seven popular SZZ algorithms are compared. Particularly, besides the vulnerability-specific variant V-SZZ \cite{bao2022v}, we consider five classic SZZ algorithms: B-SZZ \cite{bszz}, AG-SZZ \cite{agszz}, L-SZZ \cite{LRSZZ}, R-SZZ \cite{LRSZZ}, MA-SZZ \cite{MASZZ}, and the state-of-the-art LLM4SZZ \cite{LLM4SZZ}.

\noindent\textbf{Metrics.}
Consistent with prior SZZ work \cite{DBLP:conf/icse/RosaPSTBLO21,TCSZZ}, we adopt Precision, Recall, and F1-score as our evaluation metrics:
\begin{equation}
\begin{aligned}
    Precision & = \frac{|correct_c \cap identified_c|}{|correct_c|} \\
    Recall & = \frac{|correct_c \cap identified_c|}{|identified_c|} \\
    F1-score & = 2 \times \frac{Recall \times Precision}{Recall + Precision} \\
\end{aligned}
\end{equation}
where $|correct_c|$ represents the total number of true VICs in the dataset, while $|identified_c|$ denotes the number of commits identified as VICs. F1-score is the harmonic mean of Recall and Precision.

\noindent\textbf{Implementation.}
All experiments were conducted on a server with an Intel(R) Core(TM) i9-12900k @3.90GHz and 256 GB of RAM. We uniformly access LLM APIs via OpenRouter\footnote{\url{https://openrouter.ai}}. Hash references (e.g., \code{Fixes:} tags) are removed from all provided commit messages to prevent LLM cheating.

\begin{table}[!t]
\centering
\tabcolsep=3pt
\caption{Results of \T using different LLMs.}
\label{tab:model}
 \resizebox{\columnwidth}{!} {
\begin{tabular}{@{}lccccccccc@{}}  
\toprule
\multirow{2}{*}{\textbf{LLM}} & \multicolumn{3}{c}{\textbf{V-SZZ-c}} & 
\multicolumn{3}{c}{\textbf{V-SZZ-j}} &
\multicolumn{3}{c}{\textbf{Java-SZZ}}\\
\cmidrule(lr){2-4} \cmidrule(lr){5-7} \cmidrule(lr){8-10} & 
\textbf{Pre} & \textbf{Re} & \textbf{F1} & \textbf{Pre} & \textbf{Re} & \textbf{F1} & \textbf{Pre} & \textbf{Re} & \textbf{F1} \\
\midrule
GPT-4o-mini       & 0.56 & 0.57 & 0.56
                  & 0.52 & 0.40 & 0.45
                  & 0.29 & 0.42 & 0.34\\
GPT-5-mini        & 0.66 & 0.74 & 0.71
                  & 0.58 & 0.54 & 0.56 
                  & 0.28 & 0.43 & 0.34\\
Gemini-3-flash    & 0.62 & 0.59 & 0.60
                  & 0.62 & 0.43 & 0.51 
                  & 0.29 & 0.30 & 0.29\\
\midrule
GPT-5.1           & 0.66 & 0.75 & 0.71
                  & 0.58 & 0.50 & 0.54
                  & 0.31 & \textbf{0.47} & 0.37\\
Gemini-3.1-pro    & \textbf{0.73} & \textbf{0.75} & \textbf{0.74}
                  & \textbf{0.67} & 0.51 & 0.58
                  & 0.33 & 0.45 & \textbf{0.38}\\
Claude-sonnet-4-6 & 0.67 & 0.67 & 0.67
                  & 0.59 & 0.50 & 0.54
                  & 0.31 & 0.48 & 0.37\\
\midrule
DeepSeek-V3.2     & 0.63 & 0.70 & 0.67
                  & 0.61 & \textbf{0.55} & 0.58
                  & 0.30 & 0.46 & 0.37\\
Qwen3.5-plus      & 0.69 & 0.75 & 0.72
                  & 0.64 & 0.54 & \textbf{0.59}
                  & \textbf{0.39} & 0.32 & 0.35\\
\bottomrule
\end{tabular}
}
\end{table}

\subsection{RQ1: Backbone Selection}
\noindent\textbf{Setup.}
We evaluate the performance of different LLM backbones on our \T. Specifically, we select a series of popular open/closed-source LLMs, including GPT-4o-mini, GPT-5-mini, Gemini-3-flash, GPT-5.1, Gemini-3.1-pro, Claude-sonnet-4-6, DeepSeek-V3.2, and Qwen3.5-plus, covering lightweight and heavyweight models. Following Bao et al. \cite{bao2022v}, we split the V-SZZ dataset into C and Java subsets to analyze language-specific performance. This results in three datasets: V-SZZ-c, V-SZZ-j, and Java-SZZ. The comparative results for these models are shown in Table \ref{tab:model}. The best value in each column is bold.

\noindent\textbf{Results.}
Overall, heavyweight LLMs outperform lightweight ones across datasets in most cases. On the C/C++ dataset, Gemini-3.1-pro demonstrates superior performance, achieving a precision of 0.73, a recall of 0.75, and an F1-score of 0.74, corresponding to relative improvements of 4.23\%, 3.57\%, and 11.76\% on F1-score over the best-performing GPT-5-mini. On the other two Java datasets, all top-tier heavyweight LLMs exhibit similar performance.

\subsection{RQ2: Performance Comparison}
\begin{table}[!t]
\centering
\tabcolsep=3pt
\caption{Comparison with different SZZ algorithms.}
\label{tab:total}
\resizebox{\columnwidth}{!} {
\begin{tabular}{@{}lccccccccc@{}}
\toprule
\multirow{2}{*}{\textbf{Approach}} & \multicolumn{3}{c}{\textbf{V-SZZ-c}} & 
\multicolumn{3}{c}{\textbf{V-SZZ-j}} &
\multicolumn{3}{c}{\textbf{Java-SZZ}}\\
\cmidrule(lr){2-4} \cmidrule(lr){5-7} \cmidrule(lr){8-10} & 
\textbf{Pre} & \textbf{Re} & \textbf{F1} & \textbf{Pre} & \textbf{Re} & \textbf{F1} & \textbf{Pre} & \textbf{Re} & \textbf{F1} \\
\midrule
B-SZZ    & 0.67 & \underline{0.69} & 0.68
         & 0.52 & \underline{0.63} & \underline{0.57}
         & 0.07 & \underline{0.44} & 0.13\\
AG-SZZ   & 0.59 & 0.52 & 0.55
         & 0.52 & 0.48 & 0.50
         & 0.06 & 0.30 & 0.10\\
L-SZZ    & 0.70 & 0.46 & 0.56
         & 0.58 & 0.32 & 0.41 
         & 0.18 & 0.14 & 0.16\\
R-SZZ    & 0.70 & 0.46 & 0.56
         & 0.54 & 0.31 & 0.39
         & 0.15 & 0.12 & 0.13\\
MA-SZZ   & 0.56 & 0.50 & 0.53
         & 0.48 & 0.48 & 0.48
         & 0.06 & 0.30 & 0.10\\
V-SZZ    & \textbf{0.95} & 0.57 & \underline{0.71}
         & 0.41 & \textbf{0.83} & 0.55
         & 0.14 & 0.30 & 0.19\\
LLM4SZZ  & 0.72 & 0.40 & 0.51 
         & \textbf{0.73} & 0.23 & 0.35
         & \textbf{0.38} & 0.16 & \underline{0.23}\\
\midrule
\textbf{\T}   & \underline{0.73} & \textbf{0.75} & \textbf{0.74}
              & \underline{0.67} & 0.51 & \textbf{0.58}
              & \underline{0.33} & \textbf{0.45} & \textbf{0.38}\\
\midrule
\quad \emph{w/o AS}  & 0.67 & 0.75 & 0.71                    & 0.42 & 0.48 & 0.45
                  & 0.32 & 0.42 & 0.36\\
\quad \emph{w/o AB}  & 0.47 & 0.49 & 0.48
                  & 0.44 & 0.51 & 0.48
                  & 0.30 & 0.41 & 0.35\\
\bottomrule
\end{tabular}
}
\end{table}

\noindent\textbf{Setup.}
We compare the performance of \T, which employs the best-performing LLM backbone Gemini-3.1-pro in RQ1, against baselines. The comparison results are presented in Table \ref{tab:total}. The best value in each column is bold and the second-best is underlined.

\noindent\textbf{Results.}
It is observed that \T outperforms all the baselines on the three datasets in terms of recall and F1-score, with the exception of recall on V-SZZ-j. Specifically, \T achieves 4.22\%, 1.75\%, and 65.22\% relative improvement in F1-score over the best-performing baseline on the three datasets, respectively. When considering all the three evaluation metrics (nine combinations cases altogether) regarding the three datasets, \T achieves the top-2 best performance in eight out of the nine cases. Overall, \T makes a great trade-off between precision and recall. 

\subsection{RQ3: Ablation Study}

\noindent\textbf{Setup.}
To evaluate the contribution of different components, we design two ablated variants of \T by removing one key component at a time: (\ding{182}) without \textbf{A}nchor \textbf{S}election (denoted as \emph{w/o AS}): treating all patch hunks in VFCs uniformly for backtracking, without distinguishing between vulnerability-relevant and irrelevant code diffs; and (\ding{183}) without \textbf{A}gentic \textbf{B}acktracking (denoted as \emph{w/o AB}): disabling the \emph{Tracer} agent, which directly identifies the earliest commit that modified the root cause line as the VIC.

\noindent\textbf{Results.}
The last two rows present the results of our ablation study. We can observe that, across all datasets, both two key designs are essential to achieve the best performance. Particularly, without the intent-oriented anchor statement selection, the precision on the three datasets drops by 8.96\%, 59.52\%, and 3.12\%, respectively. It demonstrates the necessity of untangle vulnerability-irrelevant code hunks for more accurate VIC identification. In addition, we find that the agent-driven iterative commit backtracking module makes the greatest contribution. Compared to the vanilla variant which identifies the earliest commit like V-SZZ, \T improves the F1-score by 54.17\%, 20.83\%, and 8.57\% on the three datasets, respectively. This result underscores the potential of autonomous exploration over the commit history.

\section{Conclusion}\label{sec:conclude}
This paper presents \T, a multi-agentic SZZ algorithm for vulnerability-inducing commit identification. Given a CVE description and its corresponding fixing commit, \T summarizes the root cause of the vulnerability and employs a step-forward prompting strategy to localize vulnerability-related statements based on the change intent of each patch hunk. These vulnerable statements serve as anchors from which \T autonomously traces backward through the repository's history to find the commit that first introduced the vulnerability. Extensive experiments show that \T outperforms the state-of-the-art baselines across datasets and programming languages, achieving F1-score gains of up to 65.22\% over the best-performing SZZ algorithm.

\section*{Data Availability}
Our experimental materials can be publicly accessed at \url{https://github.com/NJUPT-SecBrain/MAS-SZZ}.

\balance
\bibliographystyle{ACM-Reference-Format}
\bibliography{sample-base}

\end{document}